\documentstyle[12pt]{article}
\def\beq{\begin{equation}}
\def\eeq{\end{equation}}
\def\bea{\begin{eqnarray}}
\def\eea{\end{eqnarray}}
\def\bq{\begin{quote}}
\def\eq{\end{quote}}

\begin{document}
\pagestyle{empty}
\begin{flushright}
\end{flushright}
\vspace*{5mm}
\begin{center}
{\bf  THE YANG-MILLS STRING AS THE $A$-MODEL ON THE TWISTOR SPACE OF 
THE COMPLEX TWO-DIMENSIONAL PROJECTIVE SPACE WITH FLUXES AND WILSON LOOPS:
THE BETA FUNCTION}
\\
\vspace*{1cm}
{\bf M. Bochicchio} \\
\vspace*{0.5cm}
Physical Jefferson Laboratory, Harvard University, Cambridge MA 02138 USA  \\
and \\
INFN Sezione di Roma  \footnote{permanent address}\\
Dipartimento di Fisica, Universita' di Roma `La Sapienza' \\
Piazzale Aldo Moro 2 , 00185 Roma  \\
e-mail: marco.bochicchio@roma1.infn.it \\
\vspace*{2cm}
{\bf ABSTRACT  } \\
\end{center}
\vspace*{5mm}
\noindent
We argue that the string theory dual to a certain sector of 
the four-dimensional Yang-Mills theory 
at large-$N$ is the $A$-model wrapping $N$ Lagrangian $D$-branes on the twistor space of the 
complex two-dimensional projective space, with a certain flux and Wilson loop background,
by finding that the target-space beta function of this $A$-model coincides   
with the large-$N$ beta function of the Yang-Mills theory.
The beta function is obtained by the quantization of the $A$-model 
Chern-Simons effective action in target space, provided a certain $B$-field
and Wilson loop background is coupled to the $A$-model world-sheet sigma-model.
The $B$-field provides the embedding of the four-dimensional space-time into the two-dimensional
base of the Lagrangian-twistorial Chern-Simons by the non-commutative large-$N$ Eguchi-Kawai reduction.
In fact, in presence of the Wilson loop and $B$-field background, the large-$N$ loop equation of the twistorial
Chern-Simons theory  implies the non-commutative vortex equation of (anti-)self-dual type of the pure Yang-Mills theory
restricted to a Lagrangian submanifold in space-time, that in turn have been shown to occur, on the Yang-Mills side,
in the localization of the loop equation of the full Yang-Mills theory in the (anti-)self-dual variables for a certain
diagonal embedding of quasi $BPS$ Wilson loops. Remarkably, the Wilsonean large-$N$ beta 
function of full Yang-Mills and the large-$N$ $A$-model target-space beta function coincide.
\vspace*{1cm}
\begin{flushleft}
\end{flushleft}
\phantom{ }
\vfill
\eject
\setcounter{page}{1}
\pagestyle{plain}

\section{Introduction}

This paper arose combining the conjecture \cite{V1}, about the link between a twistorial topological
$A$-model and (supersymmetric) Yang-Mills ($YM$) theory, with the computation of the beta function of the pure $YM$
theory via the homological localization of the loop equation \cite{MB1,MB2}. \par
The main result of this paper is that a certain version of the $A$-model defined on the twistor space
of $CP^2$, $TW(CP^2)$,\footnote{Twistor space can be defined for every orientable four-manifold, essentially as the bundle of all
the almost complex structures over the given manifold \cite{Hit}. This more general definition extends the original construction of twistor
space of hyper-Kahler manifolds. See also \cite{X,D} for a convenient mathematical description and \cite{To} for
physical applications.}
is described in target space by an effective action that has the same (Wilsonean) beta function
of pure $YM$ theory at large $N$. \par
The basic idea is that there is a version of the $A$-model on twistor-space whose target-space
equation of motion coincides with the vortex equation of self-dual ($SD$) type obtained by the localization
of the loop equation of large-$N$ $YM$ for certain quasi $BPS$ Wilson loops \cite{MB1,MB2}. \par
We recall here in the introduction the framework in which the conjecture in \cite{V1} and the 
present work originated.   
In \cite{W2} it was discovered that the scattering amplitudes of $ \cal{N}$ $=4$ $SUSY$ $YM$ at weak coupling 
can be computed in terms of the vertex operators of the topological $B$-model on a supersymmetric version, $CP^{3|4}$, of 
the twistor space $CP^3$ of the four-sphere $S^4$ \cite{Hit,X}, that is the conformal compactification of the space-time $R^4$.
The supersymmetric version, $CP^{3|4}$, of the twistor space $CP^3$, is crucial
in order to satisfy the Calabi-Yau condition, necessary for the quantum consistency of the $B$-model \cite{W2}.
As a consequence the $B$-model construction is strongly rigid and essentially applies uniquely
to the $ \cal{N}$ $=4$ $SUSY$ $YM$  theory. \par
In the $B$-model description of \cite{W2} scattering amplitudes arise by the contributions
of certain $D1$-branes of the $B$-model. 
In \cite{V1} it was conjectured that there is a $S$-dual description of the $B$-model 
theory, given by a twistorial $A$-model on the same Calabi-Yau super-manifold,
in which the scattering amplitudes may arise by means of the world-sheet instantons of the $A$-model
instead of the $D1$-branes of the $B$-model.
The occurrence of the $A$-model
was advocated, among other reasons, because of its less rigidity for deformations
with respect to the $B$-model \cite{V1}. Indeed the $A$-model does not need, for its quantum consistency,
to be defined on a Calabi-Yau manifold.
In fact it was later conjectured in \cite{V2} that
a version of the $A$-model on the non-supersymmetric twistor space $CP^3$,
obtained coupling the $A$-model to certain charged operators, would be equivalent
to the full sector of non-supersymmetric $YM$ theory in the large-$N$ limit. \par
As far as the scattering amplitudes of the $ \cal{N}$ $=4$ $SUSY$ $YM$ theory are concerned, 
the conjecture in \cite{V1} encounters the difficulty that the cohomology ring of the $A$-model
on twistor space, contrary to the one of the $B$-model, does not seem to support continuous fields,
in particular the twistor image of plane waves. Therefore the $A$-model does not seem to have room enough to
define scattering amplitudes.
This difficulty was pointed out already in \cite{V1} and in \cite{W2} and in fact it occurs
also in the non-supersymmetric version proposed in \cite{V2}. \par
In this paper we suggest a partial solution of this difficulty, in the case of the pure $YM$ theory,
by means of the Eguchi-Kawai ($EK$) large-$N$ reduction \cite{EK,Neu,Twc,DN}. Indeed the $EK$ reduction allows us to reabsorb
some continuous translational degrees of freedom in the internal color space at $ N=\infty $ (for a quick
review see \cite{Mak}).
Thus we argue that the cohomology of the $A$-model at $ N=\infty $ may support the translations and therefore some continuous field.
Technically the $EK$ reduction is realized in the $A$-model by a background $B$-field,
that implies a non-commutative gauge theory in target space \cite{SW}, for which the large-$N$
$EK$ reduction applies.
However, in this paper we do not attempt to define any $A$-model $S$-matrix for the large-$N$ $YM$ theory,
but, as a first step, we limit ourselves to the computation of the beta function of the $A$-model effective action. \par
We outline now some physical motivations for our construction.
The open strings of the $A$-model end on a Lagrangian submanifold \cite{W1}. Thus the Lagrangian submanifold
supports a gauge theory in target space. If $N$ $D$-branes wrap around the Lagrangian submanifold, the gauge theory
has gauge group $U(N)$. This is taken into account, in the world-sheet language, by Chan-Paton factors is some representation
of $U(N)$. We choose the fundamental representation. As it is well known \cite{W1} the $A$-model effective action is a $U(N)$
Chern-Simons ($CS$) theory defined on a Lagrangian submanifold, modified by the insertion of Wilson loops, 
that arise from the world-sheet instantons corrections of the $A$-model.
A Lagrangian submanifold in twistor space is three-dimensional, so that the twistor projection to the
space-time base is two-dimensional. 
Thus the effective $CS$ gauge theory contains only a two-dimensional information about the physical space-time.
The only way to recover the four-dimensional information is
by means of a large-$N$ partial Eguchi-Kawai reduction from four to two dimensions. This reduction is obtained
in the large-$N$ limit by making two of the four coordinates non-commutative in target space.
Indeed it is known that the limit of infinite non-commutativity coincides with the large-$N$ limit \cite{Mak}.
On the world-sheet this is equivalent to coupling the sigma-model to a certain $B$-field \cite{SW}. Notice that
the target space of the sigma-model, the twistor space, has real dimension six, so that the twistor
projection has real dimension four. Since the base of the Lagrangian submanifold has dimension two, only a partial $EK$ reduction
is necessary. Hence, on the sigma-model world-sheet, a  
$B$-field is needed, whose components are non-vanishing only along some real two-cycle of the four-dimensional base of the twistor fibration. 
In our construction the two-dimensional base of the Lagrangian submanifold, contrary to the two-cycle over which
the $B$-field lives, has to be embedded,
as most as possible, "diagonally" in the four-dimensional space-time.
This has to do with the special sector of observables to which our $A$-model is supposed to be related on the $YM$ side
in the large-$N$ limit.
These observables are quasi $BPS$ Wilson loops introduced in the $YM$ theory in \cite{MB1,MB2}.
These quasi $BPS$ Wilson loops are the analogs of supersymmetric Wilson loops
in theories with extended supersymmetry and are described in sect.2.
One of their features is that they are supported on a topological sphere "diagonally"
embedded in four-dimensional space-time. In fact this "diagonally" embedded sphere turns out to be the double cover of the base 
of the Lagrangian submanifold in twistor space. 
For these quasi $BPS$ Wilson loops of the pure $YM$ theory some localization properties hold \cite{MB1}
that are analogue of those of their supersymmetric cousins \cite{P}. In particular they are localized
on certain vortices equations of the non-commutative $EK$ reduction. \par
It is precisely at this stage that the conjecture in \cite{V1} plays a role in this paper.
The world-sheet instantons of the $A$-model advocated in \cite{V1} are related, via their contribution to
the $CS$ effective action, to the vortices that arise in the localization
of the loop equation for the quasi $BPS$ Wilson loops in \cite{MB1}. In turn these vortices are essential to get the
$YM$ beta function as explained in sect.2. \par
The localization in \cite{MB1} on the $YM$ side arises by homological methods, as opposed to the usual 
cohomological localization that occurs in twisted supersymmetric gauge theories \cite{P}.
The need of homological methods, to get localization on the $YM$ side, is due to the lack of supersymmetry of the pure $YM$ theory
and to the employment of a new form \cite{MB1} of the loop equation \cite{MM}, that has of course a non-local nature. 
In this new form of the loop equation a key role is played by the conformal anomaly, somehow in analogy \cite{MB2}
with the role that the chiral anomaly plays in the loop equation for the chiral ring of $ \cal{N}$ $=1 $ $SUSY$
gauge theories \cite{DV,SW2}. 
Yet, on the $A$-model side, we get of course a cohomological theory. Therefore,
loosely speaking, we may consider the cohomological localization \cite{W1}
of our proposed twistorial topological $A$-model as the counterpart, obtained by means of gauge fields/string duality \cite{W1}, 
of the homological localization that occurs in the quasi $BPS$ sector
of the large-$N$ pure $YM$ theory. \par 
The plan of the paper is as follows.
In sect.2 we describe the $YM$ side of our correspondence and we recall the localization
of quasi-$BPS$  Wilson loops and the computation of the large-$N$ $YM$ beta function, to make 
this paper as self-contained as possible. 
In sect.3 we describe the geometry of our $A$-model from a world-sheet 
perspective. In particular we define explicitly $TW(CP^2)$ and we recall that admits an almost complex structure with
vanishing first Chern class. We describe also our version, on the Lagrangian submanifold of twistor space,
of the Penrose-Ward twistor construction (see \cite{T}) of solutions of the $SD$ $YM$ equations, that is essential to our approach.
In sect.4 we recall the target-space effective action of the $A$-model 
and we compute, in our twistorial case, its beta function, finding agreement with the $YM$ computation.
In sect.5 we recall our conclusions.

\section{Quasi $BPS$ Wilson loops and beta function on the Yang-Mills side}

In this section we summarize the localization argument in \cite{MB1} to which, for reasons of space, we refer for more
details and references.
In \cite{MB1} it was found that in the large-$N$ limit there exists a sector of Wilson loops of the pure $YM$ theory,
referred to as the quasi $BPS$ sector, that possesses special non-renormalization properties, i.e. no perimeter divergence and no cusp
anomaly for backtracking cusps, somehow analogue to the sector of $BPS$ (i.e. supersymmetric) Wilson loops of theories
with extended supersymmetry.
The quasi $BPS$ Wilson loops of the pure $YM$ theory are defined on the basis of the analogy
with the following supersymmetric Wilson loops of theories with extended supersymmetry:
\bea
Tr\Psi(BPS)=Tr P \exp i\int_{C} (A_{a} dx_{a}(s)+i\phi_{b} dy_{b}(s))
\eea
satisfying  the local $BPS$ constraint:
\bea
\sum_{a} \dot x^2_{a}(s)-\sum_{b} \dot y^2_{b}(s)=0
\eea
Since scalar fields are absent in the pure $YM$ theory, it has been proposed in \cite{MB1, MB2}
that the correct analog of Eq.(1) is :
\bea
Tr \Psi (quasi-BPS)=Tr P \exp i \int_{C}( A_z dz + A_{\bar z} d{\bar z}+ D_{u} du + D_{\bar u} d{\bar u} )
\eea
($D_u=\partial_{u}+iA_{u}$ is the covariant derivative along the $u$ direction)
with the quasi $BPS$ condition:
\bea
dz(s)=du(s) \nonumber \\
d\bar z(s)=d\bar u(s)
\eea
Because of the quasi $BPS$ condition
this Wilson loop can also be interpreted as the holonomy of the non-hermitean connection:
 \bea
A+D=(A_z+D_u) dz+(A_{\bar z}+ D_{\bar u}) d \bar z
\eea
defined on a Riemann surface diagonally embedded in four-dimensional non-commutative space-time,
$ R^2 \times R^2_{\theta} $, in the limit of infinite non-commutativity, that is indeed equivalent to the commutative
large-$N$ limit \cite{Mak}. In \cite{MB1, MB2}
$A+D$ was denoted by $B$. Here, to agree with a universal convention, we reserve the symbol $B$
for the field of the $EK$ reduction that in the thermodynamic limit coincides with the background $B$-field of the $A$-model sigma-model.
Notice that the definition in Eq.(5) makes sense also in the case that the two $R^2$ factors have not the same non-commutativity,
as it is in our case.
However, in this case, the "diagonal" embedding needs to be interpreted as follows.
There is a partial Eguchi-Kawai reduction \cite{Mak} from four to two dimensions.
In such reduction the $u$ dependence of the fields is in fact
reabsorbed into the internal color space, while the fields remain $z$ dependent.
Technically this can be obtained representing the non-commutative 
coordinates and derivatives as creation and annihilation operators acting on the internal
color (Hilbert) space \cite{Mak,DN}. \par
The curvature of the two-dimensional non-hermitean connection $A+D$ is a linear combination
of only the $ASD$ part of the four-dimensional curvature.
All the considerations in \cite{MB1} apply in fact also to Wilson loops
defined by the conjugate diagonal embedding:
\bea
dz(s)=d\bar u(s) \nonumber \\
d\bar z(s)=du(s)
\eea
that corresponds to quasi $BPS$ Wilson loops whose connection is given by:
\bea
A+ \bar D=(A_z+D_{\bar u}) dz+(A_{\bar z}+ D_{ u}) d \bar z
\eea
and whose curvature is of $SD$ type. The conjugate diagonal embedding is important in this paper because
it is Lagrangian in the four-dimensional space-time and it can be lifted to a Lagrangian embedding in twistor space.
This is not the case, instead, for the diagonal embedding, since it is not Lagrangian in space-time. \par
It was argued in \cite{MB1} that the quasi $BPS$ Wilson loops have no perimeter divergence and no
cuspidal anomaly (for backtracking cusps) in the limit $B \rightarrow 0$ that coincides with the large-$N$ limit \cite{Mak}.
In addition it easy to see that they are in fact trivial to the lowest order
in perturbation theory in this limit, because of the cancellation between the propagator of $A_z$ and of $A_u$, due to the factor
of $i$ that occurs in front of $A_u$ in the covariant derivative.
This cancellation is the pure-$YM$ analogue of the cancellation that occurs for $BPS$ Wilson loops in $SUSY$ gauge theories,
between the gauge-fields and the scalar propagators, due to the factor of $i$ in front of the scalar field in Eq.(1) and to the $BPS$ condition
in Eq.(2). It is not known if the quasi $BPS$ Wilson loops of large-$N$ $YM$ are trivial
to all orders in perturbation theory or if they are non-trivial non-perturbatively. \par
The homological localization of quasi $BPS$ Wilson loops involves the use of the zig-zag symmetry of the Wilson loops and a new
form of the loop equation, that is obtained by changing variables from the connection to the $ASD$ part of the curvature. \par
The zig-zag symmetry is exploited drawing backtracking strings that start at a marked point of the loop and end into cusps at infinity.
Adding a backtracking string does not change the holonomy class of a Wilson loop, in the same way adding a co-boundary
does not change the cohomology class of a closed differential form. This, together with the fact that in the large-$N$ limit 
quasi $BPS$ Wilson loops have 
no cusp anomaly for backtracking cusps and no perimeter divergence, is the staring point of the homological 
localization pursued in \cite{MB1}. \par
Essentially the zig-zag symmetry is used to show that the right hand side of the loop equation vanishes at a backtracking cusp.
Then the left hand side reduces to the critical equation for an effective action.
To complete the argument every marked point of the loop must be mapped into a cusp at infinity.
This is achieved by a local conformal transformation on the surface over which the Wilson loop lives after introducing a lattice
regularization.
This two-dimensional conformal change lifts to a four-dimensional conformal transformation,
because of the diagonal (or conjugate diagonal) embedding of the quasi $BPS$ Wilson loop in space-time. Thus the effective action changes by the
conformal anomaly,
that is a local counter-term and thus does not change the renormalization group flow, but only the position of the
subtraction point \cite{MB1}. \par
To pursue the analogy with the usual cohomological localization, the invariance of the renormalization group flow for a transformation
that adds to the loop a backtracking cusp is 
the analog of the property that the action is a closed form. Indeed it can 
be also described as the property that the action is invariant under the transformation that generates a co-boundary, since the action
is annihilated by a $BRST$ differential in the cohomological case. Its homological counterpart is the invariance
of the flow of the renormalized effective action for the addition of a boundary, in the sense that adding to a quasi $BPS$
Wilson loop a backtracking string can be interpreted as a local change of the conformal structure of the surface over which
the Wilson loop lives. \par 
The main technical point, to get the new loop equation, is a change of variables in the $YM$ functional integral from the connection
to the $ASD$ part of the curvature. We start with the large-$N$ $YM$ theory defined on  $R^2 \times R_{\theta}^2$  in the limit of infinite non-commutativity
$\theta$, that is known to reproduce the ordinary commutative large-$N$ limit:
\bea
Z=\int \exp(-\frac{N}{2g^2} \sum_{\alpha \neq \beta} \int Tr_f (F_{\alpha \beta}^2) d^4x) DA \nonumber \\
=\int \exp(-\frac{N 8 \pi^2 }{g^2} Q-\frac{N}{4g^2} \sum_{\alpha \neq \beta} \int Tr_f(F^{-2}_{\alpha \beta}) d^4x) DA
\eea
In the second line the classical action is conveniently rewritten as the sum of a topological and a purely $ASD$ term.
The topological term $Q$ is the second Chern class, given by:
\bea
Q=\frac{1}{16 \pi^2} \sum_{\alpha \neq \beta} \int Tr_f (F_{\alpha \beta} \tilde F_{\alpha \beta}) d^4x
\eea
while the $ASD$ curvature $F^-_{\alpha \beta}$ is defined by:
\bea
F^-_{\alpha \beta}=F_{\alpha \beta}- \tilde F_{\alpha \beta} \nonumber \\
\tilde F_{\alpha \beta} = \frac{1}{2} \epsilon_{\alpha \beta \gamma \delta} F_{\alpha \beta}
\eea
Introducing the projectors, $P^-$ and $P^+$, the curvature can be decomposed into its
$ASD$ and $SD$  components:
\bea
F_{\alpha \beta}=P^-F_{\alpha \beta}+P^+ F_{\alpha \beta} \nonumber \\
= \frac{1}{2}F^-_{\alpha \beta}+\frac{1}{2}F^+_{\alpha \beta}
\eea
We change variables from the connection to the $ASD$ curvature,
introducing in the functional integral the appropriate resolution of identity:
\bea
1= \int \delta(F^{-}_{\alpha \beta}-\mu^{-}_{\alpha \beta}) D\mu^{-}_{\alpha \beta}
\eea
The partition function thus becomes:
\bea
Z=\int \exp(-\frac{N 8 \pi^2 }{g^2} Q-\frac{N}{4g^2} \sum_{\alpha \neq \beta} \int Tr(\mu^{-2}_{\alpha \beta}) d^4x)
\nonumber \\
\times \delta(F^{-}_{\alpha \beta}-\mu^{-}_{\alpha \beta}) D\mu^{-}_{\alpha \beta} DA
\eea
We can write the partition function in the new form:
\bea
Z=\int \exp(-\frac{N 8 \pi^2 }{g^2} Q-\frac{N}{4g^2} \sum_{\alpha \neq \beta} \int Tr(\mu^{-2}_{\alpha \beta}) d^4x)
\nonumber \\
\times Det'^{-\frac{1}{2}}(-\Delta_A \delta_{\alpha \beta} + D_{\alpha} D_{\beta} +i ad_{\mu^-_{\alpha \beta}} )
D\mu^{-}_{\alpha \beta}
\eea
where the integral over the gauge connection of the delta-function has been now explicitly performed:
\bea
\int DA_{\alpha} \delta(F^-_{\alpha \beta}- \mu ^-_{\alpha \beta})= |Det'^{-1}(P^- d_A \wedge)|
\nonumber \\
=Det'^{-\frac{1}{2}} ((P^- d_A \wedge)^*(P^- d_A \wedge)) \nonumber \\
=Det'^{-\frac{1}{2}}(-\Delta_A \delta_{\alpha \beta} + D_{\alpha} D_{\beta} +i ad_{ F^-_{\alpha \beta}} )
\eea
and, by an abuse of notation,
the connection $A$ in the determinants denotes the solution of the equation
$F^-_{\alpha \beta}- \mu ^-_{\alpha \beta}=0$.
The $ ' $ superscript requires projecting away from the determinants the zero modes due to gauge invariance, since
gauge fixing is not yet implied, though it may be understood
if we like to. \par
We refer to the determinant in the preceding equation as to the localization determinant,
because it arises localizing the
gauge connection on a given level, $\mu ^-_{\alpha \beta}$, of the $ASD$ curvature.
We can interpret the $ASD$ relations:
\bea
F^-_{\alpha \beta}- \mu ^-_{\alpha \beta}=0
\eea
as an equation for the curvature of the
non-Hermitean connection $A+D=(A_z+D_u) dz+(A_{\bar z}+ D_{\bar u}) d \bar z$ and a harmonic condition for the Higgs field
$\Psi=-iD=-i( D_u dz+D_{\bar u} d \bar z )= \psi+\bar \psi$:
\bea
F_{A+D} - \mu=0 \nonumber \\
\bar F_{A+D} - \bar \mu =0 \nonumber \\
d^*_A \Psi - \nu =0
\eea
that can also be written as:
\bea
F_{A+D} - \mu=0 \nonumber \\
\bar \partial_A \psi- n=0 \nonumber \\
\partial_{A} \bar \psi- \bar n=0
\eea
where the fields $\mu, \nu, n$  are suitable linear combinations
of the $ASD$ components $\mu ^-_{\alpha \beta}$.
The resolution of identity in the functional integral
then reads:
\bea
1=\int \delta(F_{A+D} - \mu) \delta(\bar \partial_A \psi- n)  \delta(\partial_{A} \bar \psi- \bar n)
D \mu Dn D \bar n
\eea
where the measure $D \mu$ is interpreted in the sense of holomorphic matrix models 
employed in the study of the chiral ring of $ \cal {N}$ $=1$ $SUSY$ gauge theories \cite{DV,SW2}. \par
The holomorphic gauge is defined as the change of variables for the connection
$A+D$, in which the curvature of $A+D$ is given by the field $\mu'$,
obtained from the equation:
\bea
F_{A+D} - \mu=0
\eea
by means of a complexified gauge transformation $G(x;A+D)$ that puts
$A+D=b+ \bar b$ in the gauge $\bar b=0$:
\bea
\bar{\partial}b_z=-i\frac{\mu'}{2}
\eea
where $\mu'=G \mu G^{-1}$.
Employing Eq.(19) as a resolution of identity in the functional integral,
the partition function  becomes:
\bea
Z=\int
 \delta(F_{A+D} - \mu) \delta(\bar \partial_A \psi- n)  \delta(\partial_{A} \bar \psi- \bar n)
 \exp(-\frac{N}{2g^2}S_{YM}) \nonumber \\
 \times \frac{D \mu}{D \mu'}
Db D \bar b  D \mu' Dn D \bar n
\eea
The integral over $(b, \bar b)$ is the same as the integral over the four $A_{\alpha}$. The resulting functional
determinants, together with the Jacobian of the change of variables to the
holomorphic gauge, are absorbed into the definition of $\Gamma$. \par
$\Gamma$ plays here the role of a classical action, since it must be
still integrated over the fields $\mu', n, \bar n$.
$\Gamma$ is given by:
\bea
\Gamma=\frac{N 8 \pi^2 }{g^2} Q
+\frac{N}{g^2} \int Tr_f(F^{-2}_{01}+F^{-2}_{02}+F^{-2}_{03} ) d^4x \nonumber \\
+ log Det'^{-\frac{1}{2}}(-\Delta_A \delta_{\alpha \beta} + D_{\alpha} D_{\beta} +i ad_{\mu^-_{\alpha \beta}})
-log\frac{D \mu}{D \mu'}
\eea
with:
\bea
\mu^0=F^-_{01}   \nonumber \\
n+\bar n=F^-_{02} \nonumber \\
i(n-\bar n)=F^-_{03}
\eea
Although $\Gamma$ is the classical action in the $ASD$ variables it contains already quantum 
corrections because of the Jacobian of the change of variables. It turns out that its divergent part
coincides with the divergent part of the Wilsonean
localized quantum effective action, after the inclusion of zero modes.
Until now the theory is still four-dimensional.
Taking functional derivatives with respect to the $ASD$ field we get, for a planar quasi $BPS$ loop,
the loop equation:
\bea
0=\int  D\mu'  Tr \frac{\delta}{\delta \mu'(w,0)}
  (\exp(- \Gamma)
  \Psi(x,x;b)) \nonumber \\
 = \int  D\mu' \exp(-\Gamma)
  (Tr(\frac{\delta \Gamma}{\delta \mu'(w,0)} \Psi(x,x;b))
  \nonumber \\
  -\int_{C(x,x)} dy_z \frac{1}{2}  \delta^{(2)}(0) \bar{\partial}^{-1}(w-y)
  Tr(\lambda^a \Psi(x,y;b)
  \lambda^a \Psi(y,x;b)) ) \nonumber \\
 =\int D\mu' \exp(-\Gamma)
  (Tr(\frac{\delta \Gamma}{\delta \mu'(w,0)} \Psi(x,x;b)) \nonumber \\
  - \int_{C(x,x)} dy_z \frac{1}{2} \delta^{(2)}(0) \bar{\partial}^{-1}(w-y)(Tr( \Psi(x,y;b))
    Tr(\Psi(y,x;b))  \nonumber \\
  - \frac{1}{N} Tr( \Psi(x,y;b) \Psi(y,x;b))))
\eea
where in our notation we have omitted the integrations $D n D \bar n$, since they are irrelevant
in the loop equation, because the curvature of $A+D$ depends only on $\mu$.
In the large-$N$ limit it reduces to:
\bea
 \tau(\frac{\delta \Gamma}{\delta \mu'(w.0)} \Psi(x,x;b)) = \nonumber \\
  \int_{C(x,x)} dy_z \frac{1}{2} \delta^{(2)}(0) \bar{\partial}^{-1}(w-y) \tau( \Psi(x,y;b))
    \tau(\Psi(y,x;b)))
\eea
Now it is natural to perform a partial $EK$ reduction from four to two dimensions.
Let us describe what in fact the partial $EK$ reduction means in this context.
We already observed that we can absorb the translations into a gauge transformation
in the four-dimensional non-commutative theory along the two non-commutative directions \cite{DN}.
As a result the classical action looks two-dimensional, in the sense that the space-time dependence
of the fields
is two-dimensional, despite the fact that the theory is truly four-dimensional.
The four-dimensional information is hidden in the central extension  $B=\frac{2\pi}{\theta}$
that shows up in the curvature due to non-commutativity \cite{DN}.
Now we use the fact that in the non-commutative theory the integral over the non-commutative directions can be represented
as a (color) trace \cite{DN}:
\bea
\int d^2u =\frac{2\pi}{B} Tr
\eea
Hence the non-commutative classical action of the reduced theory gets a volume factor of
$V_2=2\pi \theta=\frac{2\pi}{B}$ because of the gauge choice.
The equation of motion of the $EK$ reduced theory is therefore multiplied by this volume factor.
We can divide both sides of the loop equation by this volume factor in the reduced theory
in such a way that the equation of motion is normalized as in the four-dimensional theory.
Then the inverse volume will appear in the right hand side instead of the factor $\delta^{(2)}(0)$ .
We can compensate this fact by rescaling the classical action by a factor of $N_2^{-1}$ \cite{Mak,Kaw},
with $N_2=V_2 \delta^{(2)}(0)$, in such a way that the factor of $\frac{V_2}{N_2}$ in the reduced classical action
produces the factor of $\delta^{(2)}(0)=
\frac{N_2}{V_2}$ once carried to the right hand side of the loop equation. 
Of course in all this discussion we are implicitly assuming that the trace of the reduced theory
includes now the non-commutative degrees of freedom.\par
Thus the classical action of the $EK$ reduced theory in the $ASD$ variables
is given by:
\bea
\Gamma=\frac{N 8 \pi^2 }{N_2 g^2} Q
+\frac{N}{g^2} \frac{2\pi}{N_2 B}\int Tr_f(F^{-2}_{01}+F^{-2}_{02}+F^{-2}_{03} ) d^2x \nonumber \\
+ log Det'^{-\frac{1}{2}}(-\Delta_A \delta_{\alpha \beta} + D_{\alpha} D_{\beta} +i ad_{\mu^-_{\alpha \beta}})
-log\frac{D \mu}{D \mu'}
\eea
where the trace in the functional determinants has to be interpreted
coherently with the partial $EK$ reduction. \par
The new loop equation is then:
\bea
\tau(\frac{\delta \Gamma}{\delta \mu'(w)} \Psi(x,x;b)) = \nonumber \\
   \int_{C(x,x)} dy_z \frac{1}{2} \bar{\partial}^{-1}(w-y) \tau( \Psi(x,y;b))
   \tau(\Psi(y,x;b)))
\eea
This new loop equation, that is a version of the usual Makeenko-Migdal loop equation \cite{MM,MM1}, holds for the $YM$ theory after the $EK$ reduction
from four to two dimensions in the holomorphic
gauge in which the connection $A+D$ is gauge equivalent to $b$ and its curvature is gauge equivalent to $\mu'$.
After this reduction it is convenient to perform a conformal compactification
in such a way that the theory now is defined over a two-sphere $S^2$, in order to get a nice moduli problem for the gauge fields.
Before the $EK$ reduction, in the four-dimensional Euclidean theory, this would amount to a compactification
from $R^4$ to $S^2 \times S^2$. In the four-dimensional theory in ultra-hyperbolic $(2,2)$
signature the conformal compactification would amount instead to compactify to $\frac{S^2 \times S^2}{Z_2}$
where the $Z_2$ acts by the antipodal map on both $S^2$. After the $EK$ reduction the theory is thus defined on $S^2/Z_2$
in Minkowskian $(2,2)$ signature. 
The conformal compactification adds to the effective action at most a finite conformal anomaly, that can be ignored. \par
It is clear that the contour integration in the right hand side of the loop equation
includes the pole of the Cauchy
kernel. We need therefore a gauge invariant regularization.
The natural choice consists in analytically continuing the loop equation
from Euclidean to Minkowskian space-time (with ultra-hyperbolic signature, if we must keep the gauge group to be $U(N)$
in the $ASD$ equations). Thus $z \rightarrow i(x_+ + i \epsilon)$.
This regularization has the great virtue of being manifestly gauge invariant.
In addition this regularization is not loop dependent. \par
The result of the $i \epsilon$ regularization of the Cauchy kernel is the sum of
two distributions, the principal part plus
a one-dimensional delta-function:
\bea
\frac{1}{2}\bar{\partial}^{-1}(w_x -y_x +i\epsilon)= (2 \pi)^{-1} (P(w_x -y_x)^{-1}
- i \pi \delta(w_x -y_x))
\eea
The loop equation thus regularized looks like:
\bea
\tau(\frac{\delta \Gamma}{\delta \mu'(w)} \Psi(x,x;b))= \nonumber \\
\int_{C(x,x)} dy_x(2 \pi )^{-1} (P(w_x -y_x)^{-1}
 - i \pi \delta(w_x -y_x)) \nonumber \\
\times \tau( \Psi(x,y;b)) \tau(\Psi(y,x;b)))
\eea
The right hand side of the loop equation
contains now two contributions.
A delta-like one-dimensional contact term, that is supported on closed
loops and a principal part distribution that is supported 
on open loops. Since by gauge invariance it is consistent to assume
that the expectation value of open loops vanishes, the principal part 
does not contribute and the loop equation reduces to:  
\bea
\tau(\frac{\delta \Gamma}{\delta \mu'(w)} \Psi(x,x;b))= \nonumber \\
\int_{C(x,x)} dy_{x}
\frac {i}{2}\delta(w_x -y_x) \tau(\Psi(x,y;b)) \tau(\Psi(y,x;b)))
\eea
Taking $w=x$ and using the transformation properties of the holonomy of $b$
and of $\mu(x)'$, the preceding equation can be rewritten in terms
of the connection, $A+D$, and the curvature, $\mu$:
\bea
\tau(\frac{\delta \Gamma}{\delta \mu(x)} \Psi(x,x;A+D))= \nonumber \\
\int_{C(x,x)} dy_{x}
\frac {i}{2}\delta(x_x -y_x) \tau( \Psi(x,y;A+D)) \tau(\Psi(y,x;A+D)))
\eea
where we have used the condition that the trace of open loops vanishes
to substitute the $b$ holonomy with the $A+D$ holonomy. \par
We need a lattice version of the continuum loop equation
to implement our localization argument.
Thus we write the loop equation in the $ASD$ variables on a lattice in the partially $EK$ reduced theory.
If we introduce a lattice, the delta-functional constraint in Eq.(16-19) becomes, after the partial $EK$ reduction:
\bea
& [ D_{z}, D_{\bar z}]- [D_u , D_{\bar u}]- i\sum_{p} \mu^0_p \delta^{(2)}(x-x_p)+i B1=0 \nonumber \\ 
& [ D_{\bar z }, D_ u ]- i\sum_{p}  n_p \delta^{(2)}(x-x_p)=0 \nonumber \\
& [D_{z} , D_{\bar u}]-i\sum_{p}  \bar n_p \delta^{(2)}(x-x_p)=0
\eea
This system of equations defines a central extension of parabolic Higgs bundles on a sphere, in which the role of the
Higgs field $\Psi$ is played by $-iD$. Parabolic Higgs bundles have been introduced also in \cite{W4}, in their study
of the ramified Langlands conjecture.
Since the Higgs field,$-iD$, acts on the infinite dimensional Hilbert space of a non-commutative $R^2$,
the curvature equation involves a central term, $B=\frac{2 \pi}{\theta}$,
that we have displayed explicitly. 
In the case $n_p=\bar n_p=0$, that is the most relevant for us, we may interpret the preceding
equations as vortex equations.
In the partially $EK$ reduced theory, the vortices live at the lattice points, where the $ASD$ curvature
is singular. This means that in the original four-dimensional theory they form two-dimensional
vortex sheets. Codimension-two singularities of this kind occur also in \cite{W4}, but without the non-splitting
central extension in the curvature. 
The  loop equation on our lattice now reads:
\bea
\tau(\frac{\delta \Gamma}{ \delta \mu(x_p)}\Psi(x_p,x_p;A+D))=  \nonumber \\
\int_{C(x_p,x_p)} dy_z \frac{1}{2}\bar{\partial}^{-1}(x_p-y) \tau(\Psi(x_p,y;A+D))
\tau (\Psi(y,x_p;A+D)))
\eea
and correspondingly for the analytic continuation to Minkowskian space-time:
\bea
\tau(\frac{\delta \Gamma}{\delta \mu'(w_{x_q})} \Psi(x_q,x_q;A+D))=  \nonumber \\
 \int_{C(x_q,x_q)} dy_x(2 \pi )^{-1} (P(w_{x_q} -y_x)^{-1} 
  - i \pi \delta(w_{x_q} -y_x))  \nonumber \\
 \times  \tau( \Psi(x_q,y;A+D)) \tau(\Psi(y,x_q;A+D)))
\eea
Notice that a smooth marked point gives a non-trivial contribution in the continuum or on the lattice in the right hand side.
However around a backtracking cusp the contributions of the two sides of the asymptotes to the cusp
cancel each other for the contact term:
\bea
\int_{C(x_q,x_q)} dy_x(s)
\delta(w_{x_q}(s_{cusp}) -y_x(s))=\frac{1}{2} (\frac{\dot w_{x_q}(s^+_{cusp})}{ |\dot w_{x_q}(s^+_{cusp})|}+
\frac{\dot w_{x_q}(s^-_{cusp})}{|\dot w_{x_q}(s^-_{cusp})|}) 
\eea
because of the opposite sign of $ \dot w_{x_q}(s^+_{cusp})$ and $ \dot w_{x_q}(s^-_{cusp})$ 
on the two sides of the backtracking cusp.
For the principal part the same argument applies 
because of the opposite orientations of the asymptotes and because 
both the cusp asymptotes are approached either from below or from above:
\bea
|\int_{C(x_q,x_q)} dy_x(s)
P(w_{x_q}(s_{cusp}) -y_x(s))^{-1}|= \nonumber \\
\int ds \frac{1}{2} (\frac{\dot y_x(s^+_{cusp})}{|w_{x_q}(s_{cusp}) -y_x(s)|}+
\frac{\dot y_x(s^-_{cusp})}{|w_{x_q}(s_{cusp}) -y_x(s)|}) 
\eea
Thus if every marked point can be transformed into a backtracking cusp we can complete our argument
about localization, since then the loop equation reduces to the equation of motion for the
effective action in the left hand side.
But this is precisely the effect of our lattice, since marked points contribute to
the loop equation in the lattice theory only if they coincide with the lattice points.
Thus we can simply draw backtracking strings from the loop to the lattice points and then map conformally the lattice points
to infinity
in order to transform all the marked points into cusps, for which the right hand side of the loop equation vanishes. \par
Doing so we change $\Gamma$ by the conformal anomaly, into $\Gamma_q$, the quantum effective action.
Thus we may say that open strings solve 
the $YM$ loop equation for the quasi $BPS$ Wilson loops, in the sense that they localize the 
loop equation on a saddle-point for an effective action. \par 
As a consequence, the large-$N$ loop equation of
the full four-dimensional Yang-Mills theory in the anti-self-dual variables ($ASD$)
for the diagonally embedded quasi $BPS$ Wilson loops, whose connection is $A+D$, localizes on the moduli space of
a central extension of parabolic Higgs bundles.
In addition it was found in \cite{MB1} that the critical point in the loop equation actually corresponds to
non-commutative vortices equations, reduced to two dimensions a la Eguchi-Kawai:
\bea
 &[D_z, D_ {\bar z}]- [D_u , D_{\bar u}]= i \sum_{p} g \lambda_p g^{-1}  \delta^{(2)}(w-w_p)-i B1 \nonumber \\
 &[D_{\bar z }, D_ u]=0  \nonumber \\
 &[D_{z} , D_ {\bar u}]=0
\eea
It was found in \cite{MB1} that the vortices equations
imply the correct beta function of the Yang-Mills theory
in the large-$N$ limit \footnote{Since the integral over $\mu=\mu^0+n-\bar n$ has to be interpreted in a holomorphic sense,
a choice of the holomorphic path is needed. The hermitean path corresponding to
$n=\bar n=0 $ leads to the correct beta function.}.
The beta function is extracted from the $YM$
effective action $\Gamma_q$ in the $ASD$ variables computed on the vortices.
More precisely, the following result was found for the Wilsonean and canonical beta function.
There exists a renormalization scheme in which the large-$N$ canonical beta function of the pure $YM$ theory is given by:
\bea
\frac{\partial g_c}{\partial log \Lambda}=\frac{-\beta_0 g_c^3+
\frac {\beta_J}{4} g_c^3 \frac{\partial log Z}{\partial log \Lambda} }{1- \beta_J g_c^2 }
\eea
with:
\bea
\beta_0=\frac{1}{(4\pi)^2} \frac{11}{3} \nonumber \\
\beta_J=\frac{4}{(4\pi)^2}
\eea
where $g_c$ is the 't Hooft canonical coupling constant and $ \frac{\partial log Z}{\partial log \Lambda} $ 
is computed to all orders in the 't Hooft Wilsonean coupling constant, $g_W$, by:
\bea
\frac{\partial log Z}{\partial log \Lambda} =\frac{ \frac{1}{(4\pi)^2} \frac{10}{3} g_W^2}{1+cg_W^2}
\eea
with $c$ a scheme dependent arbitrary constant.
At the same time, the beta function for the 't Hooft Wilsonean coupling
is exactly one loop:
\bea
\frac{\partial g_W}{\partial log \Lambda}=-\beta_0 g_{W}^3
\eea
Once
the result for $ \frac{\partial log Z}{\partial log \Lambda} $ to the lowest order in the canonical
coupling 
\bea
\frac{\partial log Z}{\partial log \Lambda}=
\frac{1}{(4\pi)^2} \frac{10}{3} g_c^2 + ...
\eea
is inserted in Eq.(40), it implies the correct value of the first and
second perturbative coefficients of the beta function:
\bea
\frac{\partial g_c}{\partial log \Lambda}=
-\beta_0 g_c^3+
(\frac {\beta_J}{4} \frac{1}{(4\pi)^2} \frac{10}{3} -\beta_0 \beta_J) g_c^5 +... \nonumber \\
=-\frac{1}{(4\pi)^2}\frac{11}{3} g_c^3 + \frac{1}{(4\pi)^4} ( \frac{10}{3}
-\frac{44}{3})g_c^5 +... \nonumber \\
=-\frac{1}{(4 \pi)^2} \frac{11}{3} g_c^3 -\frac{1}{(4 \pi)^4} \frac{34}{3} g_c^5+...
\eea
which are known to be universal, i.e. scheme independent. \par
In addition it was argued in \cite{MB1} that there is a scheme in which the canonical coupling coincides with
a certain definition of the physical effective charge in the inter-quark potential.
In this scheme the beta function is given by:
\bea
\frac{\partial g_{phys}}{\partial log r}=\beta_0 \frac{g_{phys}^3}{1- \beta_J g_{phys}^2 }
\frac{log(r\Lambda_W)}{log(r\Lambda_W)- \frac{15}{121}} 
\eea
with $\Lambda_W$ the $RG$ invariant scale in the Wilsonean scheme. The preceding formula compares favorably
with numerical lattice computations for $SU(3)$. \par
For completeness we outline here briefly the computation of the Wilsonean beta function
since it is related to the computation made in sect.4 of this paper.
The effective action of vortices is given by:
\bea
\exp({-\Gamma_q})=\int d(zero-modes) \exp{(-\Gamma+ Conformal Anomaly)}  
\eea
with $\Gamma$, in the partially reduced $EK$ theory, given by Eq.(28).
The divergent part of $\Gamma$ is given by:
\bea
(\frac {N}{2g_W^2}-(2-\frac {1}{3})\frac{N}{(4 \pi)^2} log(\frac {\Lambda}{\tilde \Lambda}))
\sum_{\alpha \ne \beta} \int d^4x
Tr_f (2(\frac{1}{2}F^-_{\alpha \beta})^2)  \nonumber \\
=(\frac {N}{2g_W^2}-\frac {5}{3}\frac{N}{(4 \pi)^2} log(\frac {\Lambda}{\tilde \Lambda})) 
\sum_{\alpha \ne \beta} \int d^4x Tr_f (2(\frac{1}{2}F^-_{\alpha \beta})^2 \nonumber \\
=\frac {N}{2g_W^2} Z^{-1} \sum_{\alpha \ne \beta} \int d^4x Tr_f (2(\frac{1}{2}F^-_{\alpha \beta})^2)
\eea
where $Z^{-1}$ is given by:
\bea
Z^{-1}=1-\frac{10}{3} \frac{1}{(4 \pi)^2} g_W^2 log(\frac {\Lambda}{\tilde \Lambda})
\eea
and we have added to $g$ the underscript $_W$ to stress that our computation here refers
to the Wilsonean coupling constant.
We must add to this divergence the one due to vortices zero modes, that are the moduli of the adjoint orbit.
The zero modes divergence is due to the powers of the Pauli-Villars regulator that have to be inserted
in the integral over the zero modes. \par
For a $Z_N$ vortex of charge $k$ we get $N-k$ eigenvalues of the curvature $\lambda_p$ equal to
$\frac{2 \pi k}{N}$ and $k$ eigenvalues equal to $\frac{2 \pi (k-N)}{N}$.
The trace of the eigenvalues of the curvature in the fundamental representation is thus:
\bea
(N-k) (\frac{2 \pi k}{N})^2 + k (\frac{2 \pi (k-N)}{N})^2 
=(2 \pi)^2 \frac{k(N-k)}{N}
\eea
Each $Z_N$ vortex carries a number of zero modes 
equal to the real dimension of the adjoint orbit $ g \lambda_p g^{-1}$. However it must taken into account the fact that
the loop equation localizes only after analytic continuation to $(2,2)$ signature, since this analytic continuation
is needed to regularize the loop equation in a gauge invariant way. This imposes some global constraint
on the vortex solution. In $(2,2)$ signature the conformal compactification of
space-time is $S^2 \times S^2/Z_2$ where the $Z_2$ acts by the antipodal involution on both $S^2$. Its double cover is
$S^2 \times S^2$. Thus a vortex solution on Euclidean $S^2 \times S^2$ extends, after analytic continuation to Minkowski, to a vortex solution
on $S^2 \times S^2/Z_2$ only if vortices on $S^2 \times S^2$ come in pairs identified by the antipodal involution.
On the double cover the number of vortices is doubled, so that the action is doubled,
but the number of zero modes is not, because the vortex adjoint orbits must be pairwise identified.
Thus the number of real zero modes per vortex is one half, i.e. it is the number of complex zero modes.
Thus it is equal to the complex dimension of the orbit. We do not include zero
modes associated to translations of the vortices since their contribution is
subleading in $\frac{1}{N}$.
The complex dimension of an adjoint orbit of $U(N)$ is given by: 
\bea
dim=\frac{1}{2} (N^2-\sum_i m_i^2)
\eea
where $m_i$ are the multiplicities of the eigenvalues. 
For vortices this reduces to:
\bea
dim=\frac{1}{2} (N^2-k^2-(N-k)^2)=k(N-k)
\eea
In the $YM$ theory, in the thermodynamic limit, due to the contributions of vortices zero modes, the effective action reads:  
 \bea
\exp{(-\Gamma_q)}=  
\prod_p \exp{(- \frac{2 \pi}{N_2 B a^2} \frac{8 \pi^2 }{2 g_W^2} Z^{-1}
k_p(N-k_p)-c.c.)} \nonumber \\
\exp{( \frac{2 \pi}{N_2 B a^2} \frac{k_p(N-k_p)}{2} log(\frac{1}{Ba^2})+c.c.)} 
\eea
The renormalization of the Wilsonean coupling constant in the $YM$ theory now follows immediately
from the local part of the vortices effective action. It contains two terms. The one proportional to $\frac{5}{3}$ comes from the
functional determinants; the one proportional to $2$ from the vortices zero modes:
\bea
\frac{8 \pi^2 k(N-k)}{2 g^2_W(\tilde a)}
 = 8 \pi^2 k(N-k)( \frac{1}{2 g^2_W(a)}- \frac{1}{(4\pi)^2} (2+\frac{5}{3})
log (\frac{\tilde a}{a}))
\eea
and thus: 
\bea
\frac{1}{2 g^2_W(\tilde a)} 
 = \frac{1}{2 g^2_W(a)}- \frac{1}{(4\pi)^2} (2+\frac{5}{3})
log (\frac{\tilde a}{a})  \nonumber \\
= \frac{1}{2 g^2_W(a)}- \beta_{0}
log (\frac{\tilde a}{a}) \nonumber \\
\beta_0= \frac{1}{(4\pi)^2} \frac{11}{3}
\eea
We can reabsorb the coupling constant, $g_W$,into a redefinition of the subtraction point:
\bea
\exp (-\Gamma _q)=   \prod _p \exp (k_p(N-k_p) 8 \pi ^2 
\beta_0 \log (\frac{1} {e^{\frac{1}{ \beta _0 g_W^2}} Ba^2}))
\eea
There is another redefinition of the subtraction point when the conformal anomaly, due to the homological localization,
is added. This leads to the effective action:
\bea
\exp (-\Gamma _q)=   \prod _p \exp (k_p(N-k_p) 8 \pi ^2  \beta_0 \log( \frac{1}{e^{\frac{1}{\beta _0 g_W ^2}} N_D B a^2 })) 
\eea
It is easy to see that the critical point of the effective action corresponds to $Z_2$ vortices.
Thus in the large-$N$ limit the quasi $BPS$ Wilson loops of the pure $YM$ theory are localized on
the $Z_2$ vortices of $ASD$ type of the partial $EK$ reduction.

\section{The A-model world-sheet geometry}

Following the analogy with the construction of the twistorial $B$-model 
for the $ \cal{N}$ $=4$ $SUSY$ $YM$ theory at weak coupling in \cite{W2} and the conjectured $S$-duality to the $A$-model in \cite{V1}
and \cite{V2}, we would like to identify an $A$-model on twistor space dual to the large-$N$
pure $YM$ theory, more precisely to the restricted sector of quasi $BPS$ Wilson loops described in the previous section. \par
We use as a guide the localization result in \cite{MB1} for quasi $BPS$ Wilson loops.
We look for an $A$-model whose classical equations of motion reproduce the vortex equation of the large-$N$ $YM$ theory, on which
the quasi $BPS$ Wilson loops are localized.
In the next section we show, using the loop equation of the effective $CS$ theory, that the equivalence extends to the quantum level
and to the Wilsonean beta function in the large-$N$ limit. \par
Since we cannot show in general that all the $YM$ observables are localized on vortices,
we can reasonably hope at most that the $A$-model which we are looking for describes only the quasi $BPS$ sector of the $YM$ theory. \par
The natural candidates are the $A$-models defined on twistor space of a compactification of space-time.
On the $A$-model side the compactification is needed to have a well defined moduli problem
for the $A$-model world-sheet instantons.
The thermodynamic limit is then recovered in the decompactification limit.
Though the compactification appears at this stage as a technical device, it becomes apparent later that
it has in fact deeper reasons. \par
Twistor space can be defined for any orientable Riemannian four-manifold, as the bundle over the given manifold of all its almost 
complex structures compatible with the given Riemannian metric. This is a generalization of the original construction for a hyper-Kahler manifold
and for details we refer to the mathematical and physical literature \cite{Hit,X,D,To}. 
Natural compactifications of four-dimensional Euclidean space-time are the four-sphere $S^4$, the complex projective surface
$CP^2$, and the four-torus, $T^4$. Thus we have as candidates the corresponding twistor spaces. These candidates are
a priori on an equal footing, but in fact we argue that only twistor space of $CP^2$ meets all the following physical
requirements. \par
It is a deep result \cite{W3} that the $A$-model does not need to be defined on a Calabi-Yau for its quantum consistency
but only on an almost complex manifold \footnote{Topological $A$-models defined on almost complex manifolds are 
considered also in \cite{S}. Bi-hermitean models are studied in \cite{R,Z}. We would like to thank Alessandro Tomasiello for an enlightening discussion about this point.}.
Indeed the Calabi-Yau condition for a complex threefold ensures the vanishing of the chiral anomaly
in the $B$ model \cite{W1}.
For an $A$-model on a threefold there is not such a chiral anomaly, but rather a ghost number anomaly that affects only
the selection rule for the observables \cite{W1, Mar}. In particular, in order to couple the $A$-model to gravity on the world-sheet,
i.e. to define an $A$-model string theory,
we must require that the selection rule for $n$-point observables is genus independent, for the observables
to have  corrections to all orders in the string coupling constant as for a critical string theory \cite{W1,Mar}. This implies that the first Chern class of the
complexified tangent bundle vanishes. Yet this is considerably weaker than the Calabi-Yau condition since the
almost complex structure that is involved in the definition of
the tangent bundle need not to be actually complex, i.e. integrable.
Thus our first requirement is that the $A$-model twistor space has vanishing first Chern class. \par
The second requirement is that the $A$-model must support a $B$-field on twistor space that projects to a $B$-field
on the space-time base. This is necessary to identify the $B$-field with the one that implies the non-commutative
partial large-$N$ $EK$ reduction in physical space-time. \par
Finally, the third requirement is that the twistor space admits a Lagrangian submanifold over which
the equation of motion of the $CS$ effective theory implies the vortex equation associated to the
localization of quasi $BPS$ Wilson loops of the $YM$ theory. \par
We examine now as to whether these constraints can be satisfied for our candidate twistor spaces.
We start describing the twistor space of our compactified space-times \cite{Hit,X,D,To}.
$TW(S^4)=CP^3$, $TW(CP^2)=SU(3)/U(1) \times U(1)$, $TW(T^4)=T^4\times CP^1$.
Since none of these spaces is Ricci-flat, the corresponding topological models are not Calabi-Yau's.
In particular there is no way to define the $YM$ string as a $B$-model extending 
the construction in \cite{W2} to a non-supersymmetric version.
However if an $A$-model on super-twistor space $S$-dual to the the $B$-model exists, as conjectured in \cite{V1},
then its non-supersymmetric version should be naturally related to non-supersymmetric $YM$ theory \cite{V2}.
In fact in this paper we would like to make this conjecture more precise identifying a version of the $A$-model
that reproduces the $YM$ beta function in the large-$N$ limit.
Since to define the $A$-model we need an almost complex structure, we must discuss almost complex or complex structures
on twistor space.
In this case we have the embarrassment of richness since twistor spaces admit several different (almost) complex structures.
The (almost) complex structures that can be defined on the same twistor space may have different Chern classes and thus may define
nonequivalent $A$-models because of the different selection rules for the observables.
(Almost) complex structures on twistor spaces have been classified recently \cite{D}.
However it has been known for a long time that twistor space of self-dual manifolds always admits the two following (almost) complex structures \cite{D}.
The standard integrable complex structure, that is covariantly constant with respect to the Levi-Civita connection:
its Chern class is non-vanishing in all the three cases of our study.
The non-integrable almost complex structure obtained reversing the orientation of the complexified tangent bundle on the fibre
of the twistor fibration \cite{D,To}: its Chern class vanishes
for the twistor space of self-dual Einstein manifolds \cite{X,To}, thus in all our three cases.
Because the almost complex structure is not integrable, a $H$-flux appears as the torsion of the almost complex manifold \cite{W3}.
This torsion is associated to a non-trivial $B$-field. In addition we have the freedom to add a closed $B$ be field to the $A$-model
action without changing the flux and the topological nature of the $A$-model \cite{W3}.
While any component of the $B$-field along the fibre of the twistor fibration is allowed,
the component along the base must be the $B$-field needed for the $EK$ reduction.
In particular it must be vanishing small in the large-$N$ limit and in the infinite tension limit \cite{SW} of the topological theory \footnote
{The cohomological localization of the topological theory implies that the infinite tension limit is in fact exact.}. 
We must describe in more detail the action of the $A$-model in the case of an almost complex manifold
and the almost complex structure on twistor space.
It is one of the deep results in \cite{W3} that for an $A$-model on an almost complex manifold the topological action
contains a linear combination of the metric $g$ and of the two-form $J$ associated to the almost complex structure.
Thus $J$ plays the role of a $B$-field. More explicitly 
for the twistor space of a self-dual Einstein manifold the
metric $g_6$ and the almost complex structure $J$ are given in terms of the holomorphic vierbein as follows (we use the notation
of \cite{To}):
\bea
g_6= e^i \bar e^i \nonumber \\
J= i e^i \wedge \bar e^i \nonumber
\eea  
The flux $H=dJ$ can be computed using the following formulae \cite{X, To}:
\bea
d \left(
\begin{array}{r}
e^1  \\
e^2  \\
e^3  \\
\end{array}
\right) =
\left(
\begin{array}{rc}
 - \alpha &  \\
 & Tr(\alpha)\\
\end{array}
\right)  \wedge
\left(
\begin{array}{r}
 e^1 \\
 e^2 \\
 e^3 \\
\end{array}
\right) + \frac{1}{R}
\left(
\begin{array}{r}
 \bar e^2 \wedge \bar e^3 \\
 \bar e^3 \wedge \bar e^1 \\
 \sigma \bar e^1 \wedge \bar e^2 \\
\end{array}
\right) 
\eea
where $\alpha$ is an anti-hermitean matrix of one-forms that acts on $(e^1, e^2)$ and $R$ an overall length scale. $\sigma$ parametrizes 
the curvature of the base manifold relative to the fibre. 
These formulae show that the topological action contains a $B$-field, $J$, whose components are non-vanishing
along all the three orthogonal complex directions in twistor space. This is not what we want.
We need one complex line of the base without $B$-field and one with vanishing small $B$-field, to achieve the partial $EK$ reduction. 
For $\sigma=1$ twistor space is a Kahler manifold \cite{Hit,X,To}. This corresponds to the case of  $S^4$ and $CP^2$ with the integrable complex
structure.
For $\sigma=2$ twistor space of $S^4$ and $CP^2$ is a nearly Kahler manifold with the non-integrable almost complex structure \cite{X,To} \footnote{Indeed the same topological
twistor space admits two different metrics and (almost) complex structures.}.
In the case of $S^4$ and $CP^2$ with $\sigma=2$ we obtain the required $EK$ field by adding to the topological action the following 
$B$-field with vanishing torsion:
\bea
-ie^1 \wedge \bar e^1 \nonumber-i(1-\epsilon)e^1 \wedge \bar e^1 \nonumber +i(1-\frac{1}{2}\epsilon)e^3 \wedge \bar e^3 \nonumber
\eea
that cancels the $B$-field along the complex direction $1$ and creates a small $B$-field along the complex direction $2$,
at the price of creating a large compensating $B$-field along the complex direction $3$. This is not possible for $T^4$,
because it has zero $\sigma$, since
$T^4$ is a flat manifold.and thus the torsion
of the would be compensating $B$-field on the
fibre vanishes in this case.
Thus the only two suitable spaces that may lead to a stringy $A$-model 
and large-$N$ non-commutative $EK$ reduction are the four-sphere and the projective surface.
The problem with the four-sphere is that it is not a complex manifold. Thus on the four-sphere there is no notion
of the conjugate diagonal embedding $z=\bar u$, that is needed to define our Lagrangian submanifold 
in a way compatible with the embedding in space-time of quasi $BPS$ Wilson loops
\footnote{Since twistor space of the four-sphere is $CP^3$ there is a $CP^2$ inside $CP^3$ on which to define the conjugate diagonal
Lagrangian embedding. However the complex structure of this $CP^2$ does not project to a complex structure
on $S^4$ \cite{To}.}.
Only twistor space of the projective surface remains.
On the projective surface the conjugate embedding defines a Lagrangian submanifold that is isomorphic topologically to $RP^2$,
by a slight modification of the standard embedding of $RP^2$ into $CP^2$.
It lifts to a Lagrangian submanifold in twistor space that locally is $RP^2 \times RP^1$ where $RP^1$
can be taken to be a great circle in $RP^2$.
This can be explained as follows. There is a realization of $TW(CP^2)$ as the subset of $CP^2 \times \tilde CP^2$
for which $ \sum_i z_i \tilde z_i=0  $.in projective coordinates (see \cite{To} and references therein). This means that
$TW(CP^2)$ can be thought as a fibration of a complex projective surface by its orthogonal complex line, that in turn is labeled
by another complex line that belongs to the surface. Thus  $TW(CP^2)$ is a flag manifold that is a fibration
of a complex surface by a complex line that belongs to the surface \cite{To}. When restricting to the Lagrangian submanifold
the previous statements continue to hold in their real version. \par
Now we must show that on this Lagrangian submanifold the $CS$ theory at classical level leads to the 
vortex equation.
The target-space action of the $A$-model in the open string sector is the effective action: 
\bea
S = \frac{1}{g_s} \int Tr( A F - \frac{1}{3} A ^3)+ \sum_{i} \eta _i e^{-a(\gamma _i)} Tr( P \exp i \int_{ \gamma _i} A )
\eea
defined on a Lagrangian submanifold of $TW(CP^2)$. It is not restrictive to assume that $\eta _i=1$ \cite{W1}.
$e^{-a({\gamma _i})}$ is the weight, (not necessarily real in presence of a $B$-field), by which an instanton
of (complex) area $a$, whose boundary is $\gamma _i$, is weighted in the world-sheet expansion of the $A$-model.
The extra term with respect to the usual $CS$ action is due to
world-sheet instantons corrections and to the insertion of Wilson-loop operators on the world-sheet.
Notice that the effective action is obtained re-summing the world-sheet genus expansion in a non-trivial
background of Wilson loops. Thus for its existence we must assume that the ghost number anomaly selection rule
is satisfied in a genus independent way. This in particular requires that the first Chern class of the almost complex structure vanishes
as we anticipated. 
Since the connection that can be coupled to the $A$-model is flat, the conjugacy class of the Wilson loop holonomy depends
only on the homology class of $\gamma _i$. \par
We must show that a vortex equation arises as a solution of this $CS$ theory on the Lagrangian submanifold.
Motivated by the embedding of the space-time ($CP^2$) with coordinates $(w , \bar w , u ,\bar u)$
into the twistorial fibration with coordinates $( \lambda, u_1,u_2)$:
\bea
 w-\lambda \bar u=u_1 \nonumber \\
 u+\lambda \bar w=u_2
\eea
we make the following ansatz for the $CS$ connection on the twistorial fibration in terms of the gauge fields
$(A_w, A_{\bar w}, A_u, A_{\bar u})$ on the four-dimensional space-time $CP^2$. This ansatz is very well known 
as the Penrose-Ward construction and in fact
it is at the heart of the link between self-dual $YM$ equations and flat equations along the tangent direction 
to twistor surfaces \cite{T}.
Our main point is that we need a certain version of it, restricted to the
Lagrangian submanifold, in such a way that the $CS$ equations gets deformed into the vortices
equations.
Our ansatz for the covariant derivatives of the $CS$ connection along the tangent vector fields
of the double cover of the Lagrangian fibration is:
\bea
 &D_{u_1}=D_{\bar w}- \lambda D_u   \nonumber \\
 &D_{u_2}= D_{w}+\frac{1}{\lambda} D_{\bar u}
\eea
where $ D_u= \partial _u+ i A_u $.
We employ the double cover because we want to use complex coordinates.
Physically this ansatz corresponds to imposing that the $CS$ gauge fields that live in a neighborhood of the tangent directions
of the Lagrangian submanifold contain the $YM$ gauge fields 
in a way that respects the geometry of the twistor fibration. In particular on our Lagrangian submanifold
both $\lambda$ and $-\frac{1}{\lambda}$ occur, because of the antipodal identification on the fibre $RP^1$.
From the point of view of the underlying string theory defined by the $A$-model
this occurs because the gauge fields are associated to the open strings that live indeed on the Lagrangian
submanifold. However, we must check that our ansatz can be satisfied in fact by the equation
of motion of the string theory, that are the $CS$ equations of motion.
The curvature equations without the condensate are now:
\bea
 [D_{u_1} , D_{u_2}] = \frac{1}{\lambda } [D_{\bar w} , D_{\bar u}]- ( [D_w ,D_{\bar w}]+ [D_u , D_{\bar u}]) \nonumber \\
 +\lambda  [D_w , D_u]=0
\eea
Hence they are satisfied provided the self-dual curvature of the underlying $YM$ connection vanishes.
In this case the $CS$ theory would be defined on the moduli space of $SD$ connections.
We now require that the $(u, \bar u)$ coordinates become non-commutative. This corresponds to turning on a $B$-field
in the $A$-model sigma-model. This $B$-field implies a non-vanishing flux in general and it was already implicit
in our choice of an almost complex structure.
The physical reason for which we require this $B$-field is that we must convey in the $CS$ functional integral, that contains only
a two-dimensional Lagrangian submanifold of space-time,
the physical information about the four-dimensional nature of the space-time. This is not possible at finite $N$,
but it is possible at $N =\infty $, by the $EK$ reduction. This is realized by making  
space-time non-commutative in the $(u,\bar u)$ directions in the limit of infinite non-commutativity
and by reabsorbing the non-commutative degrees of freedom in the infinite dimensional (at $N= \infty $)
Hilbert color space of the $CS$ theory. Because of the limit of infinite non-commutativity the $B$-field is
vanishing small in the large-$N$ limit. \par
As a result Eq.(62) gets deformed to:
\bea 
\frac{1}{g_s} [D_{u_1}, D_{u_2}]+ i \frac{1}{g_s} B 1 =0
\eea
since now some partial derivatives are non-commutative.
Finally we include the condensate of Wilson loops. We suppose that the Wilson loop extends only along $\lambda$ and it is
based over the point $w_p$ of the double cover of $RP^2$. The double covering is needed to allow for the existence of
complex coordinates. The curvature equation now reads:
\bea 
\frac{1}{g_s} [ D_{u_1}, D_{u_2} ] ^a=-i \frac{1}{g_s} B 1^a+i \delta ^{(2)} (w-w_p)
e^{-a(\gamma _p)} Tr(P T^a \exp i \int_{\gamma_p} A_{\lambda} d \lambda)
\eea
Thus we see that the effect of instantons is to introduce a vortex singularity
in the (non-commutative) $SD$ equations. For a certain choice of the background Wilson loop and of the string coupling
these become identical to the $Z_2$ non-commutative vortex equation of $SD$ type of the $YM$ theory:
\bea
 &[D_w , D_{\bar w}]+ [D_u , D_{\bar u}]= i g \lambda _p g^{-1}  \delta^{(2)}(w-w_p)+ \frac{i}{g_s} B1  \nonumber \\
 &[D _w, D_u  ] =0   \nonumber \\
 &[D_{\bar w}, D_{\bar u}]  =0
\eea
repeated with an infinite degeneracy, $N_1$, along the fibre.
This occurs for a certain condition on the string coupling (see next section) and for the natural choice:
\bea
(\exp i \int_{\gamma_p} A_{\lambda} d \lambda)^2=1
\eea
in such a way that $A_{\lambda}$ is flat on $RP^2$. We explain now why this choice is natural.
Turning on a non-trivial Wilson loop background on the fibre implies a compatibility condition involving
the $CS$ flatness condition on the fibre and on the base (see Eq.(70)).
We will not examine in detail this compatibility condition, but we observe that
flatness conditions of this type have been studied in the mathematical literature under the name of twistor structures \cite{Moc}.
In particular solutions exist for holomorphically flat connections on the fibre, that correspond
to our ansatz, since the holonomy on the fibre is in a representation of the fundamental group
of $RP^2$, because its square is one, and thus it is flat. \par
Even if the fibre becomes non-commutative because of the background $B$-field along the fibre,
the vortex equation is satisfied on the base. Indeed then $\lambda$ becomes a hermitean operator.
The flatness condition still implies the three $SD$ equations provided $(1, \lambda, \lambda^{-1})$ are linearly independent as operators. 
The equation for the hermitean part of the $CS$ curvature is unchanged.
In the next section we study these equations at classical and quantum level as well.

\section{The $A$-model beta function via its Chern-Simons effective action}

The beta function of our twistorial $A$-model arises as follows.
The quantization of the standard $CS$ theory, without the condensate of Wilson loops, is described by the functional 
integral \cite{M}:
\bea
Z=\int \exp (\frac{1}{g_s} Tr \int  \tilde{A}  \partial_{\lambda} \tilde{A} d\lambda) 
\delta( \frac{1}{g_s} [D_{u_1}, D_{u_2}]) D\tilde{A}
\eea
where the connection one-form $\tilde{A}$, on the double cover of the base of the Lagrangian submanifold,
is given by:
\bea
\tilde{A}=(A_w+\frac{1}{\lambda}D_{\bar u})dw+(A_{\bar w}-\lambda D_{u})d\bar w 
\eea
We call the exponent in Eq.(67) the kinetic term, because it is the only term that contains derivatives along the fibre.
The delta-functional constraint imposes the flatness condition to the gauge connection tangent to the base.
We will see that, after integrating the constraint in the delta-functional and after the inclusion of the condensate
of Wilson loops, the kinetic term 
defines a quantum mechanical theory on the moduli space of the vortex equation.
We expect that such quantum mechanical system is ultraviolet finite, being a one-dimensional quantum field theory.
Thus the only possible source of divergences comes from the functional determinant 
that arises by integrating the constraint in the delta-functional.
This is the situation in which the $YM$ beta function occurs, since, although the theory looks two-dimensional on the
base, it is in fact four-dimensional, because of the $EK$ reduction. The asymmetry between the base and the fibre is
created by the presence of the Wilson loop condensate, that we choose along a non-contractible loop on the fibre,
in order to reproduce the correct vortex equation. \par
The presence of the condensate modifies the flatness constraint in a way
that we can understand as follows.
The functional integral in presence of a condensate is given by:
\bea
Z=\int \exp(\frac{1}{g_s} Tr \int  \tilde{ A} \partial_{\lambda} \tilde{A} d\lambda) \nonumber \\
\exp ( \int  \frac{1}{g_s}
Tr(A_{\lambda}[D_{u_1}, D_{u_2}])  dw \wedge d\bar w \wedge d\lambda -
  e^{-a(\gamma _p)} Tr(P \exp ( i \int_{\gamma_p} A_{\lambda} ))) DA
\eea
If the chosen background loop, $\gamma_p$, extends in the $\lambda$ direction only and is based on the 
point $w_p$ of the base, the classical equations of motion of the $CS$ effective action, as shown in the previous section, are:
\bea
F^a_{\lambda w}=0 \nonumber \\
F^a_{\lambda \bar w}=0  \nonumber \\
\frac{1}{g_s} F^a_{w \bar w}(\tilde{A}) -i \delta ^{(2)}(w-w_p) e^{- a(\gamma_p)} Tr( P T^a \exp i \int_{\gamma_p} A_{\lambda} d\lambda)=0
\eea
where $T^a$ are the generators of $U(N)$ in the fundamental representation.
Once our choice of the background Wilson loop is made, the world-sheet instantons that dominate
the world-sheet theory are constant on the base and wrap once along the fibre.
Notice that the presence of the condensate introduces a $\delta ^{(2)}(w-w_p)$ singularity in the curvature of
the $CS$ connection on the base. This $\delta ^{(2)}(w-w_p)$ singularity arises integrating, along the fibre on which
the background Wilson loop lies,
the $\delta ^{(3)}$ singularity due to the functional differentiation in three dimensions.
In addition the presence of a condensate of Wilson lines with non-trivial holonomy along the fibre 
implies a twist in the dependence of the fields of the base on the fibre coordinate. 
The dependence on the fibre in the coefficient of $\delta ^{(2)}(w-w_p)$
is through the holonomy of the connection $A_{\lambda}$.
This holonomy dependence determines the vortex moduli space and leads to the mentioned one-dimensional quantum mechanics
on the fibre, due to the kinetic term. \par 
We now show that in the large-$N$ limit the condensate equation does not get quantum corrections
to the leading order, under certain assumptions.
The standard Makeenko-Migdal loop equation \cite{MM,MM1} applied to the $CS$ theory with the condensate reads schematically:
\bea
\tau((\frac{1}{g_s} [D_{u_1},D_{u_2}](w_p)-i \delta^{(2)}(w-w_p) e^{- a(\gamma_p)} \Psi(\gamma_p ; A))\Psi(\gamma_p ; A)) \nonumber \\
 = i \delta^{2}(w-w_p) \tau(\Psi(\gamma_{p} ; A)) \tau(\Psi(\gamma_{p}; A))
\eea
where $\tau$ is a generalized trace, that is the combination of the v.e.v. with the normalized color trace and 
$\Psi(\gamma_{p}; A)=P \exp ( i \int_{\gamma_p} A_{\lambda} )$ is
the holonomy of the Wilson loop. The term that contains the double loop in the left-hand side is due to the condensate.
The term on the right hand side is the usual interacting $MM$ term. \par
If the background of Wilson loops is non-trivial, i.e. $\tau(\Psi(\gamma_{p}; A))=0$, the interaction term vanishes and 
thus the classical equation is not renormalized in the large-$N$ limit.
This is precisely the situation that is needed to reproduce the $YM$ beta function.
The $YM$ beta function is reproduced by a background of Wilson loops on the fibre that implies 
a $Z_2$ vortex in space-time and thus satisfies the non-trivial
condition $\tau(\Psi(\gamma_{p}; A))=0$.
Indeed the equation for the hermitean part of the $CS$ curvature on the double covering of the base of the Lagrangian submanifold
reads:
\bea 
-\frac{1}{g_s} ( [D_w , D_{\bar w}]+[D_u, D_{\bar u}] )(w_p) = -\frac{i}{g_s}  B +i e^{- a(\gamma_p)}\Psi(\gamma_p;A) \delta ^{(2)} (w-w_p) 
\eea
The holonomy along the fibre, $\Psi(\gamma_p;A)$, leads to the $Z_2$ vortex provided
$g_s e^{- a(\gamma_p)}\Psi(\gamma_p;A)= g \lambda_p g^{-1}$, for some $g$
unitary, is the curvature of the gauge connection of a $Z_2$ vortex.
The only obvious solution is $g_s e^{- a(\gamma_p)}=\pi$ and the eigenvalues of $\Psi(\gamma_p;A)$ equal to $(1,-1)$
in equal number. Thus $\tau(\Psi(\gamma_{p}; A))=0$. \par
Now we compute the beta function by saturating the $MM$ loop equation.
By saturating the $MM$ equation we mean writing a functional integral that satisfies the same quantum equation
of motion, i.e. the same $MM$ loop equation, to the leading $\frac{1}{N}$ order.
Hence this saturating path integral contains a delta-functional
involving the equation of motion of one vortex and its antipodal image, since we are on the double covering:
\bea
Z=\int \prod_{\lambda} \delta( F_{w \bar w}(\tilde{A}) 
-  i \delta ^{(2)} (w-w_p) 
  g \lambda _p g^{-1} - i g  \lambda _p g ^{-1} \delta ^{(2)} (w-\bar w_p))  \nonumber \\
   D\tilde A_{w}(\lambda) \wedge D \tilde A_{\bar w}(\lambda) 
\eea
with $\lambda_p$ the curvature of the $YM$ connection of a $Z_2$ vortex at $p$, that means that $\lambda_p$ 
has eigenvalues equal to $(\pi, -\pi)$ in equal number. \par 
This  functional integral is apparently two-dimensional, but since it is repeated on the fibre, it reproduces
the integral over the four-dimensional gauge connections with the same degeneracy by which the $SD$ constraint is repeated.
Hence it reproduces the four-dimensional information.
To see this, we recast it in the following form:
\bea
\int  DA_w DA_{\bar w} DA_u DA_{\bar u} \nonumber \\
 \delta([D_w , D 
 _{\bar w}]+ [D_u , D_{\bar u}]-i g \lambda_p g^{-1}  \delta^{(2)}(w-w_p)
-i g \lambda_p g^{-1}  \delta^{(2)}(w -
\bar w_p)-i B 1) \nonumber \\
\delta([D_w , D_u]) \delta([D
_{\bar w}, D_{\bar u}])  
\eea
repeated with an infinite multiplicity $N_1$, along the fibre.
In the decompactification limit
we can identify the  base of the Lagrangian twistorial fibration with space-time of $YM$ theory in
the thermodynamic limit. Thus Eq.(74) becomes:
\bea
\int \delta(F^{+}_{\alpha \beta}-\mu^{+}_{\alpha \beta}) DA   \nonumber \\
=\int d(zero-modes) Det'^{-\frac{1}{2}}(-\Delta_A \delta_{\alpha \beta} + D_{\alpha} D_{\beta} +i ad_{ F^+_{\alpha \beta}} )
\Delta_{FP} 
\eea
for 
\bea
\mu^{+}_{01}= g\lambda_p g^{-1}  \delta^{(2)}(w-w_p)+g\lambda_p g^{-1} \delta^{(2)}(w-\bar w_p)+B1 \nonumber \\
\mu^{+}_{23}= g\lambda_p g^{-1} \delta^{(2)}(w-w_p)+g\lambda_p g^{-1} \delta^{(2)}(w- \bar w_p)+B1 
\eea
and all other components of $\mu^{+}_{\alpha \beta}$ vanishing.
This functional integral, in the $EK$ reduced version, was already computed in sect.(2) and in \cite{MB1} but for the $ASD$
variables instead of the $SD$ ones. Of course this does not change the divergences.
We can compare the result for the effective action of vortices of the $YM$ theory computed in sect.3 with the $CS$
functional integral of Eq.(73,74,75).
The exponential of minus the $CS$ effective action is:
\bea
 \exp(N_1 k_p(N-k_p) 8 \pi ^2  \beta_0 \log ( \frac{ \Lambda ^{2}_{CS} }{B_{CS}} ))
\eea
with $k_p=\frac{N}{2}$ since in the $CS$ theory we get only a $Z_2$ vortex with multiplicity $N_1$.
The factor of $N_1$ is due to the multiplicity of the fibre delta-functional constraint in the $CS$ functional integral.
In \cite{MB1} it was shown that the $YM$ Wilsonean beta function is one-loop exact at large
$N$ using the homological localization of the loop equation. The $CS$ beta function shares the same feature.
Thus we have the identification $\frac{ \Lambda ^{2}_{CS} }{B_CS}= \frac{1}{N_D B e^{\frac{1}{ \beta _0 g_W ^2}} a^2 } $
that is our result.

\section{Conclusions}

We have shown that there exists an $A$-model on the twistor space of $CP^2$ with a non-integrable complex 
structure that has vanishing first Chern class. Thus it defines a topological string theory.
In addition its Chern-Simons effective action at large $N$, on a certain Lagrangian submanifold
and for a certain background Wilson loop and $B$-field, has the same Wilsonean beta function
as the large-$N$ Yang-Mills theory.
This is due to the identification of the twistor
Chern-Simons loop equation in the given background with the non-commutative vortex equation
of self-dual type of the Yang-Mills
theory reduced a la Eguchi-Kawai. \par
Conjecturally, following the analogy with the topological $B$-model of the $ \cal{N} $ $=4$ $YM$ theory,
this topological $A$-model holds promise to describe the glueball dynamics of
a certain quasi $BPS$ sector of the pure $YM$ theory. 

\section{Acknowledgments}

We thank Andrew Neitzke and Cumrun Vafa for several discussions about the $A$-model and the $YM$-loop
equation and Cumrun Vafa for inviting us to complete this work at Harvard University. \par
We thank Arthur Jaffe for inviting us to talk at the seminar of his group about the localization of the $YM$ 
loop equation and the $A$-model. \par
We thank Denis Auroux and Tomasz Mrowka at $MIT$ for explaining to us some features of the twistor fibration. \par
We thank Roberto Martinez for several stimulating conversations
about the $A$-model and the $YM$ loop equation.

\end{document}